\documentclass[epj,nopacs]{svjour}
\usepackage{graphics}
\usepackage{graphicx}
\usepackage[figuresright]{rotating}
\usepackage{epsfig}
\usepackage{subfigure}
\newcommand{\pion}{$\pi^0$ }
\newcommand{\epair}{e$^+$e$^-$--pair }
\newcommand{\epairs}{e$^+$e$^-$--pairs }
\newcommand{\pt}{$p_T$ }

\newcommand{\AuAu}{$\mathrm{Au} + \mathrm{Au}$ }

\begin{document}
\title{Measurement of photons via conversion pairs in $\sqrt{s_{NN}}$~=~200~GeV
Au$\mathbf{+}$Au collisions with the PHENIX experiment at RHIC}
\titlerunning{Measurement of photons via conversion pairs with PHENIX at RHIC}
%
\author{Torsten Dahms\inst{1} for the PHENIX collaboration
}                     
\institute{Department of Physics and Astronomy, Stony Brook
  University, Stony Brook, NY 11794-3800, USA, \\
}
\date{Received: date / Revised version: date}
%
\abstract{Thermal photons can provide information on the temperature
  of the new state of matter created at RHIC. In the \pt region of
  1--3~GeV/c thermal photons are expected to be the dominant direct
  photon source. Therefore, a possible excess compared to a pure decay
  photon signal due to a thermal photon contribution should be seen in
  the double ratio
  $(\gamma/\gamma(\pi^{0}))_\mathrm{Measured}/(\gamma/
  \gamma(\pi^{0}))_\mathrm{Simulated}$, if sufficient accuracy can be
  reached. We present a method to reconstruct direct photons by
  measuring \epairs from external photon conversions.}
\maketitle

\section{Introduction}
\label{sec:introduction}
Direct photons are produced during all stages of heavy ion collisions
at the Relativistic Heavy Ion Collider (RHIC). Because they do not
interact strongly, they escape the medium unaffected by final state
interactions and provide a promising signature of the earliest and
hottest stage of the quark-gluon plasma (QGP)~\cite{Turbide:2003si}.

On a microscopic level, the main sources of direct photons from a QGP
are quark-gluon Compton scattering ($q g \rightarrow \gamma q$),
quark-anti\-quark annihilation ($q \bar{q} \rightarrow \gamma g$) and
brems\-strahlung involving thermalized
partons~\cite{Aurenche:1998nw}. Direct photons are also produced in
initial hard scattering processes which involve the same reactions but
among the incoming particles.

At RHIC energies thermal photons are predicted to be the dominant
source of direct photons in a \pt window between 1--3
GeV/c~\cite{Turbide:2003si}.

Direct photons have been measured with PHENIX in \AuAu collisions at
$\sqrt{s_{NN}} = 200$~GeV~\cite{Adler:2005ig}. The inclusive photon
spectra measured with the Electromagnetic Calorimeter (EMC) have been
compared to the expected background from hadronic sources, based on
the measured \pion and $\eta$ spectra and a cocktail of other hadronic
decays ($\eta^{\prime}$, $K_S^0$, $\omega$), assuming $m_T$ scaling.
\begin{figure}[b]
  \centering
  \includegraphics[width=0.45\textwidth]{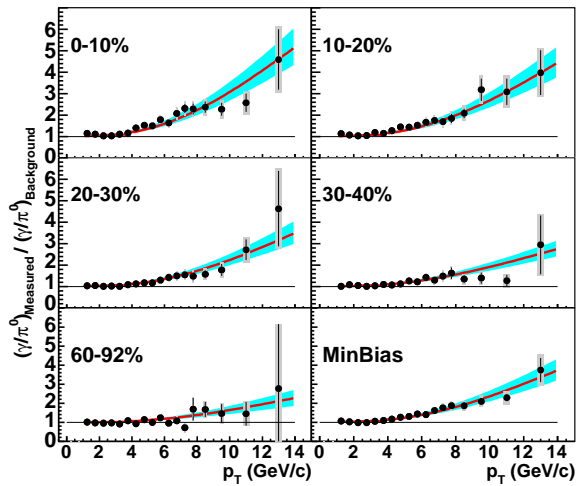}
  \caption{Double ratio of the measured invariant yield ratio,
    $\left(\gamma/\pi^0\right)_\mathrm{Measured}$, to the background
    decay ratio, $\left(\gamma/\pi^0\right)_\mathrm{Background}$, as a
    function of \pt for minimum bias and for five centralities of
    \AuAu collisions at $\sqrt{s_{NN}} = 200$~GeV.}
  \label{fig:doubleratio}
\end{figure}

Fig.~\ref{fig:doubleratio} shows the double ratio of the measured
invariant yield ratio to the background decay ratio as a function of
\pt for minimum bias and for five centrality classes. The measurement
of direct photons production at high \pt scales with the number of
binary collisions in agreement with NLO perturbative QCD predictions
and therefore confirms medium effects as the origin of jet
suppression. In the low \pt region, where a thermal signature is
expected, a significant measurement remains limited by systematic
uncertainties due to the energy resolution and the photon
identification with the EMC.

In order to overcome such limitations, dielectron pairs offer some
advantages because of the superior resolution of charged particles at
low momenta and excellent identification of conversion photons; while
other methods~\cite{Bathe:2005nz} try to use low mass dielectron pairs
from internal conversions, the method presented here uses real photon
conversions in the beam pipe.

\section{Thermal photon analysis}
\label{sec:analysis}
The excellent capabilities of the PHENIX detector to measure electrons
suggest to circumvent the limitations of the conventional direct
photon measurement~\cite{Adler:2005ig} at low photon energies by
measuring photons via their conversion pairs. The momentum resolution
($\sim 1~\%$) of the charged tracking devices proves superior to the
energy resolution of the EMC ($\sim 10~\%$) in the \pt region of
interest (1--3~GeV/c).

Two steps are used to identify \epairs from photon conversions. First
a single electron identification cut is applied, which require signals
from at least two phototubes in the Ring Imaging Cherenkov Detector
(RICH) matching to a reconstructed charged track in the Drift Chamber
(DC). No further electron identification cuts were applied since the
pair cuts (see Sect.~\ref{sec:conversions}) to separate conversion
photons from other \epairs are more efficient and powerful enough to
provide a very clean photon conversion sample.

The extracted photon conversions are tagged with photons reconstructed
in the EMC to determine the contribution from $\pi^0 \rightarrow
\gamma \gamma$ decays (see Sect.~\ref{sec:tagging}).

All yields are measured as a function of $p_T$ of the
e$^+$e$^-$--pair, which makes a direct comparison of the inclusive
photon yield, $N_{\gamma}^\mathrm{incl}$, and the tagged photon yield,
$N_{\gamma}^{\pi^0 \mathrm{tag}}$, possible:
\begin{eqnarray}\label{eq:inclusive}
N_{\gamma}^\mathrm{incl}\left(p_T\right) &=&
\epsilon_{e^+e^-}~a_{e^+e^-}~\gamma^\mathrm{incl}\left(p_T\right) \\
\label{eq:piontag}
N_{\gamma}^{\pi^0 \mathrm{tag}}\left(p_T\right) &=&
\epsilon_{e^+e^-}~a_{e^+e^-}~
\epsilon_{\gamma}\left(p_T\right)~f~\gamma^{\pi^0}\left(p_T\right)
\end{eqnarray}
The measured yield of inclusive photons depends on the reconstruction
efficiency $\epsilon_{e^+e^-}$ and the PHENIX acceptance $a_{e^+e^-}$
of the conversion \epair. The tagged photon yield depends in addition
on the efficiency to reconstruct the second photon in the EMC
$\epsilon_{\gamma}(p_T)$ and on the conditional probability $f$ to
find it in the EMC acceptance, given that the \epair has been
reconstructed already. Here $\epsilon_{\gamma}(p_T)$, as well as all
other yields and correction factors, are determined as a function of
the $p_T$ of the $e^+e^-$--pair. Therefore, in a ratio of inclusive
conversion photons to conversion photons which have been tagged as
$\pi^0$ decay products, the \epair reconstruction efficiency and
acceptance correction factor cancel.

A ratio of the hadronic decay photon yield, $N_{\gamma}^{hadr}$, and
the tagged photon yield from $\pi^0$ decays, $N_{\gamma}^{\pi^0 tag}$,
is calculated with simulations, for which again the acceptance
correction cancels.
\begin{eqnarray}\label{eq:hadronic}
N_{\gamma}^\mathrm{hadr}\left(p_T\right) &=& a_{e^+e^-}~\gamma^\mathrm{hadr}\left(p_T\right) \\\label{eq:pions}
N_{\gamma}^{\pi^0 \mathrm{tag}}\left(p_T\right) &=& a_{e^+e^-}~f
~\gamma^{\pi^0}\left(p_T\right)
\end{eqnarray}

The comparison of the ratio in data and in simulations in a double
ratio leads to an expression that is equivalent to the ratio of
inclusive and decay photons as shown in Eq.~(\ref{eq:doubleratio}). In
here also the conditional probability $f$ of finding the second photon
in the PHENIX acceptance, once the \epair is already reconstructed,
cancels.
\begin{eqnarray}\label{eq:doubleratio}
\frac{\gamma^\mathrm{incl}\left(p_T\right)}{\gamma^\mathrm{hadr}\left(p_T\right)} =
     \frac{\epsilon_{\gamma}\left(p_T\right) \cdot \left(\frac{N_{\gamma}^\mathrm{incl}\left(p_T\right)}{N_{\gamma}^{\pi^0 \mathrm{tag}}\left(p_T\right)}\right)_\mathrm{Data}}
     {\left(\frac{N_{\gamma}^\mathrm{hadr}\left(p_T\right)}{N_{\gamma}^{\pi^0 \mathrm{tag}}\left(p_T\right)}\right)_\mathrm{Sim}}
\end{eqnarray}
The only remaining factor is the reconstruction efficiency of the
photon in the EMC, $\epsilon_{\gamma}(p_T)$, which has been determined with Monte Carlo
simulations to be, independent of the pair-\pt, $95.0 \pm 1.0~\mathrm{(syst)}~\%$.

\subsection{Photon Conversions}
\label{sec:conversions}
Since the PHENIX tracking algorithm assumes the track to originate
from the collision vertex, off-vertex conversion pairs are
reconstructed with an artificial opening angle which leads to an
invariant mass that is proportional to the radius at which the
conversion occurs.

\begin{figure}[h]
  \centering
  \includegraphics[width=0.45\textwidth]{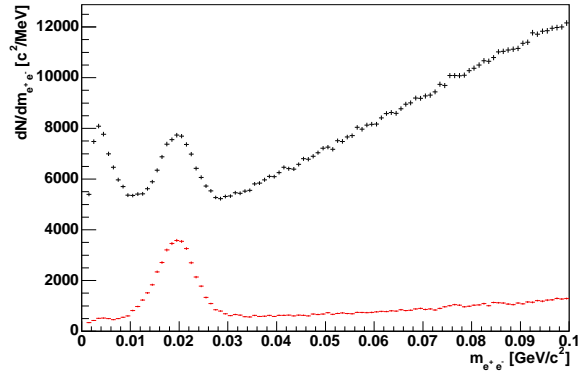}
  \caption{Invariant mass of \epairs before (black) and after (red)
    applying cuts on the orientation of the \epair in the magnetic
    field.}
  \label{fig:allpairs}
\end{figure}

Therefore, photon conversions that occur in the beam pipe material
(Be, $0.3~\%$ radiation length) at a radius of 4~cm are reconstructed
with an invariant mass of $\sim
20~\mathrm{MeV/c^2}$. Fig.~\ref{fig:allpairs} shows an invariant mass
spectrum of \epairs in the range 0--0.1~GeV/c$^2$. The peak from
photon conversions in the beam pipe at 20~MeV/c$^2$ can be clearly
separated from Dalitz decays $\pi^0 \rightarrow \gamma e^+ e^-$, which
dominate the spectrum below 10~MeV/c$^2$, and combinatorial background
pairs, whose contribution increases toward higher invariant masses.

The photon conversion pairs can be distinguished from Dalitz decays
and purely combinatorial pairs by cutting on the orientation of the
\epair in the magnetic field. The magnetic field inside PHENIX is
parallel to the beam axis. Therefore conversion pairs, which have no
intrinsic opening angle, are bent only in the azimuthal direction
(i.e. perpendicular to the direction of the magnetic field). In
contrast, the opening angle of Dalitz pairs and combinatorial pairs
can have any opening angle with respect to the magnetic
field. Furthermore, the azimuthal opening angle of conversion pairs
will always have the same sign, if one considers the ordered
difference:
\begin{equation}\label{eq:deltaphi0}
\Delta \varphi_0 = \varphi_0\left(e^-\right) - \varphi_0\left(e^+\right)
\end{equation}

\begin{figure}[t]
  \centering
  \includegraphics[width=0.45\textwidth]{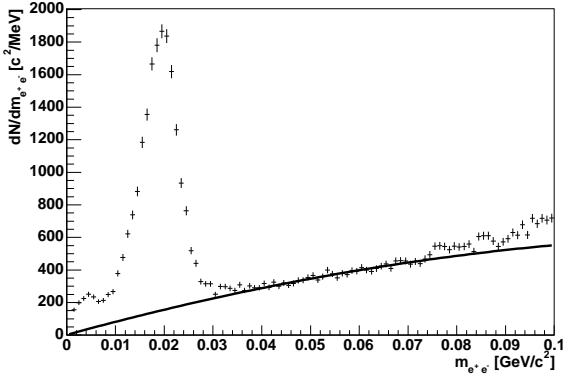}
  \caption{Invariant mass of \epairs after applying pair cuts in the
  pair-$p_T$ region 0.8--1.2~GeV/c. The black line is indicating the
  remaining contamination by combinatorial background.}
  \label{fig:convpairs}
\end{figure}

Fig.~\ref{fig:allpairs} shows the invariant mass spectra of \epairs
before (black) and after (red) applying these pair cuts. The yield
from integrating the mass region $ < 35~\mathrm{MeV/c}^2$ of the
conversion peak is corrected for the remaining $p_T$ independent
contamination of $15.0 \pm 2.0~\mathrm{(syst)}~\%$ from combinatorial
\epairs which has been determined with mixed events (see black line in
Fig.~\ref{fig:convpairs}). This leads to
$N_{\gamma}^\mathrm{incl}\left(p_T\right)$, the total yield of
conversion photons as a function of $p_T$.

\subsection{Tagging of Decay Photons}
\label{sec:tagging}
\begin{figure}
  \centering
  \includegraphics[width=0.45\textwidth]{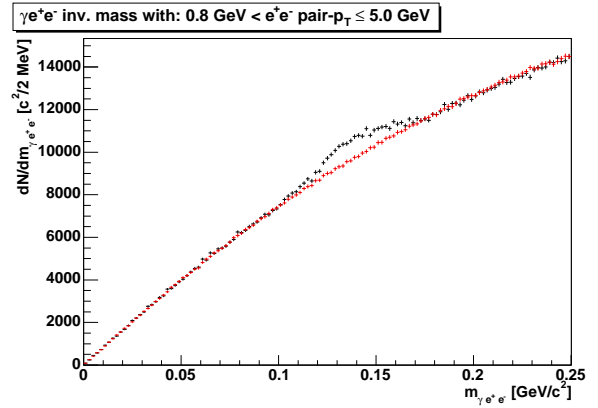}
  \caption{Invariant mass of $\gamma e^+ e^-$--triplets in same events
    (black) and normalized mixed events (red) for \epairs with
    $0.8~<~p_T~\leq~5.0$~GeV/c.}
  \label{fig:invmasstriplets}
\end{figure}

To reveal which of these conversion photons come from $\pi^0
\rightarrow \gamma \gamma$ decays, the \epairs in the conversion peak
are combined with photons which have been measured in the EMC and
their invariant mass is calculated (see
Fig.~\ref{fig:invmasstriplets}). Conversion photons that are
identified as decay products of $\pi^0$ can be tagged as
$N_{\gamma}^{\pi^0 \mathrm{tag}}$. This signal has a large
combinatorial background due to the high photon multiplicity in \AuAu
collisions.

The combinatorial background is reproduced with an event mixing
method, which creates uncorrelated pairs of photons and \epairs from
different events. The mixed event spectra is normalized to the same
event invariant mass spectra well outside the $\pi^0$ mass region and
subtracted. The normalization factor $\alpha$ is calculated as:
\begin{eqnarray}
\alpha = \frac{1}{2} \left( \frac{FG1}{BG1} + \frac{FG2}{BG2} \right)
\end{eqnarray}
with $FG1$ ($BG1$) indicating the yield in the foreground (background)
in the mass region 0--100~MeV/c$^2$ and $FG2$ ($BG2$) the
region 170--250~MeV/c$^2$.

\begin{figure}
  \centering
  \includegraphics[width=0.45\textwidth]{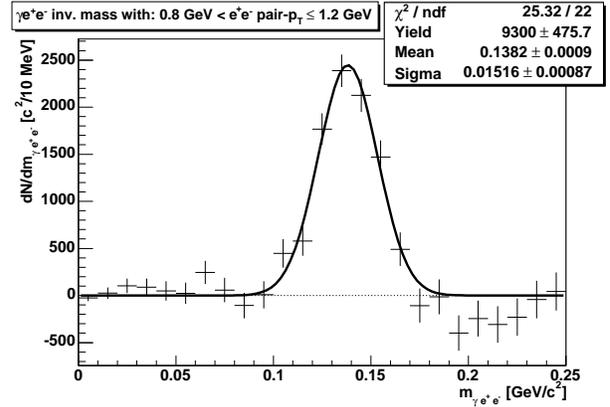}
  \caption{Invariant mass of $\gamma e^+ e^-$--triplets after
    background subtraction for \epairs with $0.8 < p_T \leq
    1.2$~GeV/c. A fit with a Gaussian is drawn and the resulting
    parameters shown in the box in the upper right of the graph.}
  \label{fig:invmasspion}
\end{figure}

The normalization factor has a relative statistical error on the order
of $0.2~\%$ which depends only on the size of the normalization region
in the foreground:
\begin{eqnarray}
\frac{\sigma_{\alpha}}{\alpha}= \sqrt{\frac{1}{FG1 + FG2}}
\end{eqnarray}
As an example, the resulting $\pi^0$ signal for \epairs with $0.8 <
p_T \leq 1.2$~GeV/c is shown in Fig.~\ref{fig:invmasspion}.

Mean and $\sigma$ are determined by a fit of the background subtracted
data with a Gaussian. For comparison the data are also fitted to the
sum of a second order polynomial and a Gaussian, to take into account
the possibility that the shape is not completely described by the
mixed event spectrum. The difference in the resulting mean and
$\sigma$ is negligible. The mean and $\sigma$ obtained by the fit are
then used to integrate the data in a region $\pm 1.5~\sigma$ around
the mean. The integration region has been chosen to optimize the
signal to background ratio.

The statistical error on the extracted $\pi^0$ signal is given by:
\begin{eqnarray}
\sigma_S^2 &=& \sum_i FG(i) + \alpha \sum_i BG^{\prime}(i) +
  \left(\frac{\sigma_{\alpha}}{\alpha} \sum_i
  BG^{\prime}(i)\right)^2\label{eq:staterror}
\end{eqnarray}
With $FG(i)$ and $BG^{\prime}(i)$ being the yields in bin $i$ of invariant
mass spectrum in same events and normalized mixed events,
respectively, the summations are performed over the integration
region. It is important to note that the last term in
Eq.~(\ref{eq:staterror}), is the square of the sum over the normalized
background, and therefore, depends on the integration region and is
not bin independent. The systematic errors on the peak extraction have
not been evaluated yet.

\begin{figure}[t]
  \centering
  \includegraphics[width=0.45\textwidth]{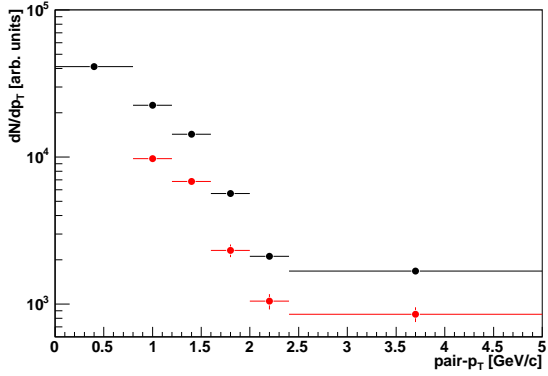}
  \caption{Number of conversion photons, $N_{\gamma}^\mathrm{incl}$
    (black), and number of tagged \pion, $N_{\gamma}^{\pi^0 \mathrm{tag}}$
    (red), as a function of \pt. (The yields are not corrected for the bin
    width in \pt since this cancels in the ratio.)}
  \label{fig:pairpt}
\end{figure}

In Fig.~\ref{fig:pairpt} the \pion yield, $N_{\gamma}^{\pi^0
\mathrm{tag}}$, as a function of pair-\pt is compared to the \pt
distribution of all conversion photons $N_{\gamma}^\mathrm{incl}$. Due
to the limited PHENIX acceptance, $\gamma e^+ e^-$ triplets can not be
reconstructed for \epairs with $p_T \leq 0.8$~GeV/c.

\subsection{Simulations}
\label{sec:simulations}
The contribution of hadronic decays has been determined with a fast
Monte Carlo simulation of $\pi^0$ and $\eta$ Dalitz decays. A
parameterization of the $\pi^0$ spectrum measured by
PHENIX~\cite{Adler:2003qi} has been used as input. The $\eta$
distribution has been generated assuming $m_T$ scaling ($p_T
\rightarrow \sqrt{p_T^2 + m_{\eta}^2 - m_{\pi^0}^2}$) of the $\pi^0$
spectral shape and a normalization at high \pt to $\eta / \pi^0 =
0.45 \pm 0.10$, which is in agreement with PHENIX
data~\cite{Adler:2006hu,Adler:2004ta}.

In the \pt region of interest, \epairs from Dalitz decays and from
photon conversions have the same shape, which allows to circumvent a
full Monte Carlo simulation of photon conversions in PHENIX, which
would be a cumbersome process due to the low conversion probability of
$0.2~\%$ in the beam pipe.

\begin{figure}
  \centering
  \includegraphics[width=0.45\textwidth]{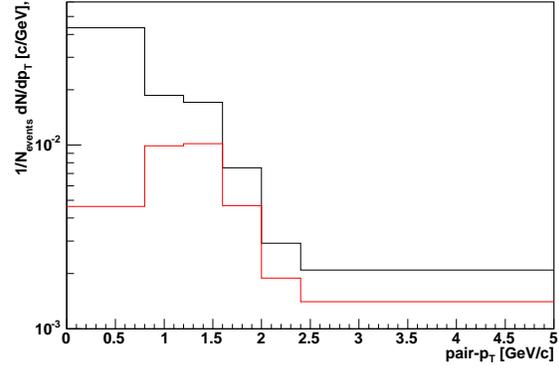}
  \caption{Shown in black is $N_{\gamma}^\mathrm{hadr}$, the number of
    \epairs from hadronic sources ($\pi^0$ and $\eta$) as a function
    of \pt from a fast Monte Carlo Simulation. The number of \epairs
    from \pion with the requirement that both, \epair and photon, were
    inside the PHENIX acceptance, $N_{\gamma}^{\pi^0 \mathrm{tag}}$, is
    shown in red.}
  \label{fig:simulation}
\end{figure}

In order to produce inclusive and tagged photon yields comparable to
the data, all particles ($e^+$, $e^-$ and $\gamma$) have been filtered
into the PHENIX acceptance. Fig.~\ref{fig:simulation} shows
$N_{\gamma}^\mathrm{hadr}$, the \pt distribution of all \epairs from
hadronic sources, and $N_{\gamma}^{\pi^0 \mathrm{tag}}$, the \epairs
which originated from $\pi^0$ decays.

The ratio in Eq.~(\ref{eq:doubleratio}),
$\gamma^\mathrm{incl}\left(p_T\right) /
\gamma^\mathrm{hadr}\left(p_T\right)$, would result from dividing the
ratio of spectra shown in Fig.~\ref{fig:pairpt},
\begin{eqnarray}
\left(N_{\gamma}^\mathrm{incl}(p_T) / N_{\gamma}^{\pi^0
\mathrm{tag}}(p_T)\right)_\mathrm{Data}\nonumber
\end{eqnarray}
corrected for the photon reconstruction efficiency, $\epsilon_{\gamma}(p_T)$, by the one from
the spectra in Fig.~\ref{fig:simulation},
\begin{eqnarray}
\left(N_{\gamma}^\mathrm{hadr}(p_T) / N_{\gamma}^{\pi^0
\mathrm{tag}}(p_T)\right)_\mathrm{Sim}.\nonumber
\end{eqnarray}
An excess above 1 would be interpreted as a signal from direct
photons.

\section{Conclusions} 
\label{sec:conclusions}
While a result can not be presented yet, the method introduced in here
seems promising to find a signature of thermal photons in the low \pt
region as an excess in the double ratio as shown in
Eq.~(\ref{eq:doubleratio}). A detailed understanding of the
combinatorial background in the invariant mass spectra of $\gamma e^+
e^-$--triplets which is currently the major source of systematic
uncertainties will reduce the systematic errors significantly.

\end{document}